# Determination of centrality in nuclear collisions by using hadron calorimeter

A.Kurepin[1] *, A.Litvinenko[2], E.Litvinenko[2]

[1]Institute for Nuclear Research RAS, Moscow

[2]Joint Institute for Nuclear Research, Dubna

**Abstract**. The determination of the centrality of nuclear collision or the value of the impact parameter of heavy nuclei is of great importance for the analysis of all experimental data and comparison with theory. One method is to measure the number of spectators using a hadron calorimeter located at a small angle to the nuclear beam. It is shown that with an achievable resolution of the hadron calorimeter in energy, the accuracy of determining the impact parameter is insufficient for using the calorimeter in the MPD / NICA and CBM / FAIR projects. The error reaches 35 % at a beam energy of 2.5 GeV even for peripheral collisions. Secondary processes during the passage of spectators through the nucleus give an additional contribution to the error for central collisions and at medium centralities.

## 1. Introduction

Despite the relativistic compression of colliding nuclei in the center-of-mass system, in the quasi classical approximation their transverse dimensions do not change even at ultrahigh energies. Obviously, the result of the interaction of nuclei, especially heavy nuclei, will depend on the geometry of the collision, i.e. on the magnitude of the impact parameter. In an non central collision, the asymmetry of the interaction leads to an azimuthal dependence of the yield of reaction products, which come out in the form of an elliptical or higher order particle flux. Comparison of experimental data with calculations by relativistic hydrodynamics led to the conclusion that the particle flux is emitted from a quark-gluon matter formed in a collision, in which there is practically no viscosity [1].

For comparison with theory, it is important to determine the coefficients of the anisotropic flow of charged particles depending on the centrality of the collision at different energies. From a rather strong dependence of the asymmetry coefficients on centrality, especially the elliptic flow coefficient ( Fig. 1), it follows that the accuracy of determining the impact parameter should be no worse than 10%.

*e-mail: kurepin@inr.ru

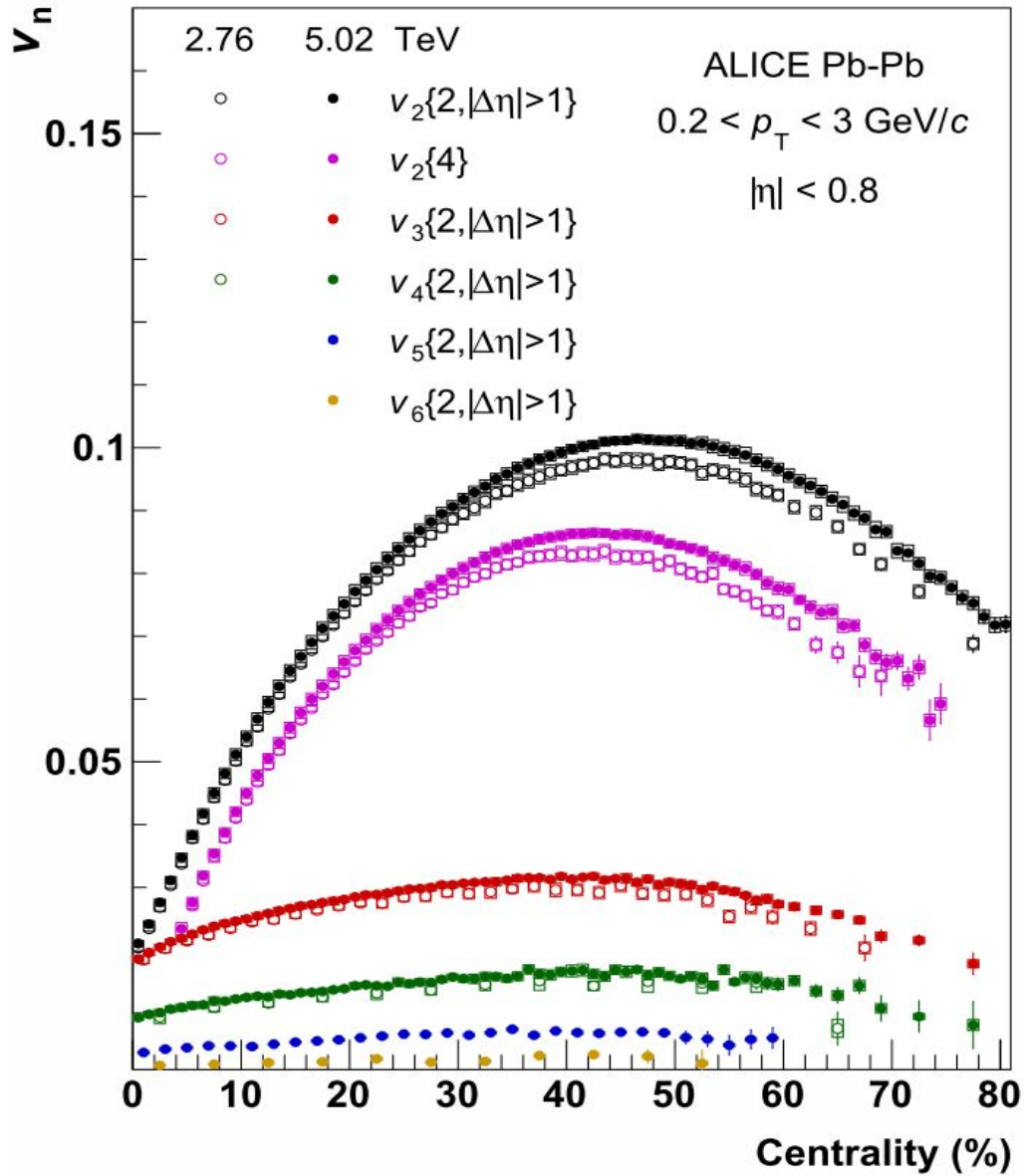

Fig. 1. Anisotropic flow coefficients for inclusive particle production depending on the centrality of the collision of lead nuclei. Solid points - energy 5.02 TeV, open points - 2.76 TeV [2]

In almost all measurements at the ALICE on the LHC, centrality was determined by measuring the amplitude of the signal of the scintillation detector V0 [3] located at high rapidity, and also by using the data of the central tracker ITS and the TPC projection camera [2,4] at rapidity < 0.8. It was shown that the total number of charged particles almost does not deviate from the proportional dependence on the number of interacting nucleons (Npart), which determine centrality [5] ( Fig. 2). Measurement of the distribution of charged particles by pseudo-rapidity at different centralities allows to study the accuracy of the predictions of various models of HIJING, EPOS, KLN, etc. [5].

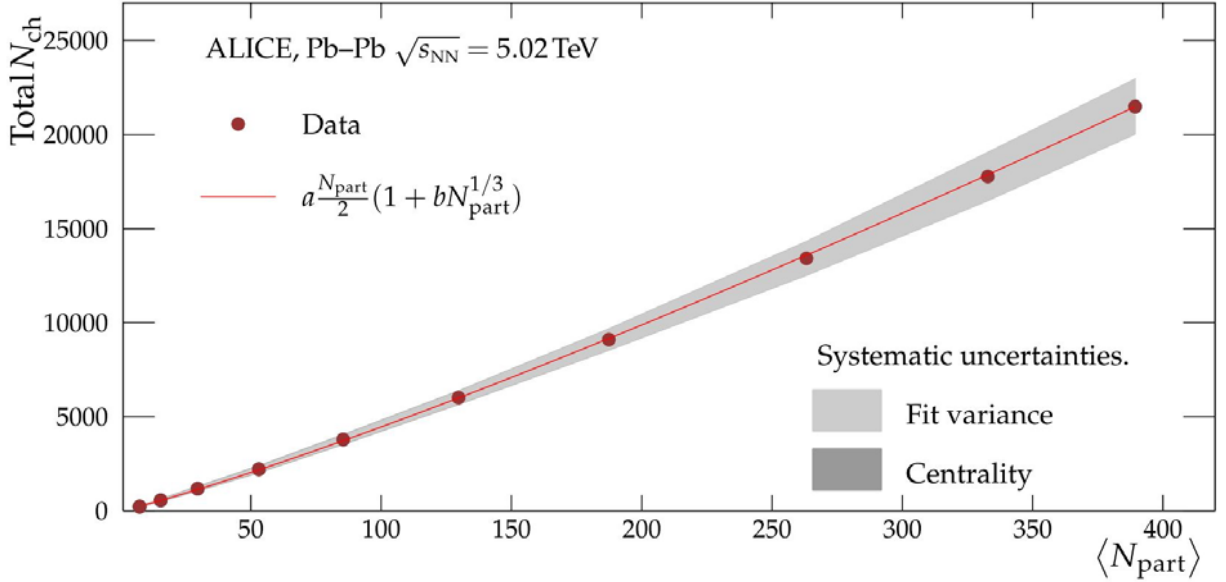

Fig. 2. Dependence of the total number of produced particles in the interval of pseudo-rapidities ($-3.5 < \eta < 5$) on the average number of interacting nucleons [5]

## 2. Determination of accuracy of the measurement of the impact parameter

To reduce the role of systematic errors in determining centrality, the use of several methods is of great importance. One of the methods for determining the centrality of the collision of heavy nuclei is the detection of spectators, i.e. particles that have not experienced interaction during a collision. Then the number of particles participating in the collision that determine centrality is equal to the difference between the number of nucleons in the nucleus and the number of spectators. This method has been successfully used at high energies at SPS, RHIC, and LHC accelerators, where spectators are recorded by hadron calorimeters placed at small angles with respect to the nuclear beam [ 3 ].

However, at lower energies this method has a number of significant features. This, in particular, is the ambiguity associated with the loss of part of the spectators leaving through the hole for the passage of the main beam, and the insufficient energy resolution of hadron calorimeters at low energies. If the ambiguity problem can be solved by measuring the asymmetry of the number of spectators, depending on the angle of emission of the spectators [ 6 ], the problem of the accuracy of determining centrality, i.e. collision parameter, requires additional consideration. The estimation of the resolution of the collision parameter for the MPD project at NICA was made in [ 7 ]. In this paper, a detailed discussion of the problem is carried out.

The hadron calorimeter was calibrated using a proton beam in a wide energy range, and it was shown that the spectrum of the calorimeter signals is described by a Gaussian curve [ 6 ]:

$$W(E)\,dE = \frac{1}{\sqrt{2\pi\sigma^2}}\,\exp\left(-\frac{(E-E_b)^2}{2\sigma^2}\right)\,dE,$$

where $E_b$ is the energy of the proton beam,

E is the calorimeter signal value, expressed in units of beam energy.

Calibration results give only for the main contribution of the stochastic term:

the variance: $\quad D = \sigma^2 = 0.31\,E_b$

the root mean square error: $\quad \sigma = 0.56\,\sqrt{E_b}$

Obviously, the distribution of absorbed energy for one spectator has the same Gaussian form with the same variance value:

$$W(E)\,dE = \frac{1}{\sqrt{2\pi\sigma^2}}\,\exp\left(-\frac{(E_s-E_b)^2}{2\sigma^2}\right)\,dE,$$

where $E_s$ is the spectator energy absorbed in the calorimeter.

When several spectators are emitted, the energy $E_t$ is released in the calorimeter:

$$E_t = E_b \cdot \bar{N}_s,$$

where $\bar{N}_s$ is the average number of spectators for a given event with the desired centrality.

However, the value of $E_t$ can correspond to the events with a different centrality, i.e. with a value of $N_s$, which differs from $\bar{N}_s$. To determine the accuracy of centrality, it is necessary to calculate the width of the distribution of $N_s$ provided that:

$$N_s \cdot E_s = E_t$$

The distribution of $E_s$ is known. However, it should be taken into account that the energy $E_t$ for a given centrality also fluctuates by the Poisson distribution with variance:

$$D(E_t) = E_b^2 \cdot \bar{N}_s$$

To find the distribution of $N_s$:

$$N_s = \frac{E_t}{E_s}$$

we use the definition of the variance of the quotient of two random variables $x$ and $y$, $E_t$ and $E_s$ for mathematically expected X and Y [ 8 ]:

$$D\left(\frac{x}{y}\right) = \frac{\sigma_x^2}{Y^2} - \frac{2X}{Y^3}\text{cov}(x,y) + \frac{X^2}{Y^4}\sigma_y^2$$

In our case for random variables $E_t$ and $E_s$ :

$$X = E_b \cdot \bar{N}_s, \quad Y = E_b, \quad \sigma_x = E_b\sqrt{\bar{N}_s}, \quad \sigma_y = \sigma$$

Since the parameters $E_t$ and $E_s$ are independent:

$$\text{cov}(E_t, E_s) = 0$$

Finally the variance and rms of $N_s$ distribution are :

$$D(N_s) = D\left(\frac{E_t}{E_s}\right) = \sigma_n^2\left(\frac{E_t}{E_s}\right)$$

$$\sigma_n = \sqrt{\left(\bar{N}_s + \frac{\bar{N}_s^2}{E_b^2}\sigma^2\right)}$$

We can compare the relative errors for a single spectator :

$$\delta_1 = \frac{\sigma}{E_b}$$

and for $N_s$ spectators:

$$\delta_2 = \frac{\sigma_n}{\bar{N}_s} = \sqrt{\left(\frac{1}{\bar{N}_s} + \frac{\sigma^2}{E_b^2}\right)} = \sqrt{\left(\frac{1}{\bar{N}_s} + \delta_1^2\right)}$$

It can be seen that the contribution of the first term under the root is significant only for the most central events. For all other events, $\delta_1 \approx \delta_2$, and the accuracy of determining centrality is determined by the energy resolution of the calorimeter for one spectator. A numerical estimate for a peripheral event at an energy of 5.5 GeV per nucleon and $\bar{N}_s = 100$ gives an estimate of the lower limit of accuracy of centrality determination of 26 %, for 2.5 GeV the accuracy is 37 %.

Estimates of the distribution of the number of spectators for given values of the collision parameter *b* are shown in Fig. 3. As a result of the secondary interaction of the spectators, a different distribution width of the number of spectators is observed at different centralities, i.e. at various values of the path of the spectator in the nucleus during the scattering or absorption. From the results of Fig. 3 it can be seen that even with an ideal resolution for the energy of the hadron calorimeter, the accuracy of determining centrality from the registration of the number of spectators cannot be better than 25% for central events and about 10% for events with an intermediate value of the collision parameter. Moreover, all the kinetic energy of the spectators is concentrated in the solid angle $5^0$.

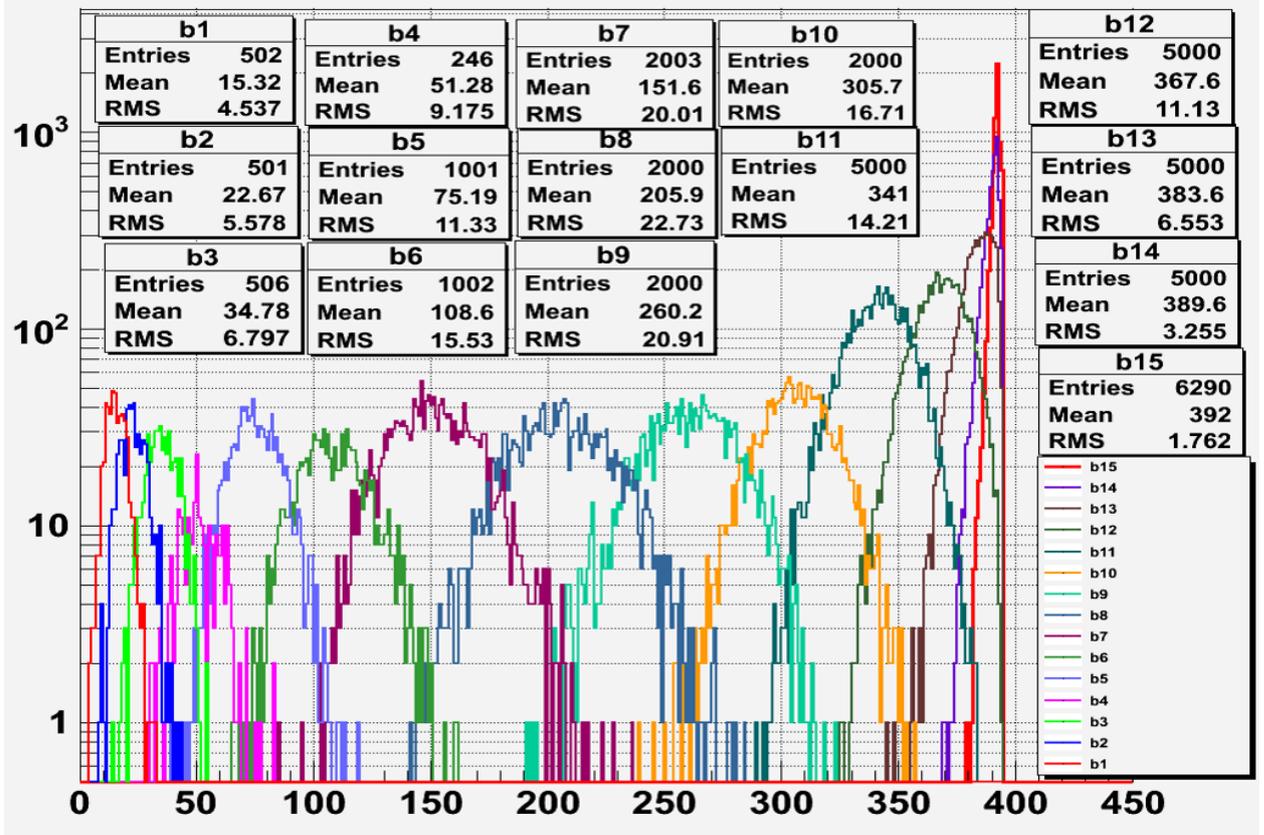

Fig. 3: Width of the distribution of the number of spectators for a fixed value of the collision parameter using the LAQGSM at the energy √s = 9 GeV

## 3. Conclusion

The results of evaluating the accuracy of determining the impact parameter or centrality of the collision coincide with the assessment of article [7]. In the entire range of centrality, the relative standard error in the determination of the impact parameter is equal or larger than the relative calibration error for single proton, which reaches 25-35% at NICA energies. Secondary processes during the passage of spectators through the nucleus give an additional contribution to the error for central collisions and at medium centralities. Since it is impossible to improve the energy resolution of the calorimeter at low energies, it is necessary to install an additional multiplicity detector, which, in combination with the hadron calorimeter, will determine the impact parameter with an accuracy of no worse than 10% over the entire range of centrality, as proposed in [7].